# Microsoft Academic Automatic Document Searches: Accuracy for Journal Articles and Suitability for Citation Analysis[1]

Mike Thelwall, Statistical Cybermetrics Research Group, University of Wolverhampton, UK.

Microsoft Academic is a free academic search engine and citation index that is similar to Google Scholar but can be automatically queried. Its data is potentially useful for bibliometric analysis if it is possible to search effectively for individual journal articles. This article compares different methods to find journal articles in its index by searching for a combination of title, authors, publication year and journal name and uses the results for the widest published correlation analysis of Microsoft Academic citation counts for journal articles so far. Based on 126,312 articles from 323 Scopus subfields in 2012, the optimal strategy to find articles with DOIs is to search for them by title and filter out those with incorrect DOIs. This finds 90% of journal articles. For articles without DOIs, the optimal strategy is to search for them by title and then filter out matches with dissimilar metadata. This finds 89% of journal articles, with an additional 1% incorrect matches. The remaining articles seem to be mainly not indexed by Microsoft Academic or indexed with a different language version of their title. From the matches, Scopus citation counts and Microsoft Academic counts have an average Spearman correlation of 0.95, with the lowest for any single field being 0.63. Thus, Microsoft Academic citation counts are almost universally equivalent to Scopus citation counts for articles that are not recent but there are national biases in the results.

## 1 Introduction

Citation-based indicators frequently support formal and informal research evaluations (Wilsdon, Allen, Belfiore, Campbell, Curry, Hill, et al. 2015). They are typically gathered from Scopus or the Web of Science (WoS), both of which index large numbers of journal articles and some other document types. Previous research has found Google Scholar to return higher citation counts than Scopus and WoS for most fields (Falagas, Pitsouni, Malietzis, & Pappas, 2008; Halevi, Moed, & Bar-Ilan, 2017) because of its inclusion of open access online publications in addition to publisher databases. It is not possible to use Google Scholar for large-scale citation analyses because it does not allow automatic data harvesting (Halevi, Moed, & Bar-Ilan, 2017), except for individual academics through the Publish or Perish software (Harzing, 2007). Microsoft Academic, which was officially released in July 2017, is like Google Scholar in its coverage of academic literature, harvesting from publishers and the open web (Harzing & Alakangas, 2017ab; Paszcza, 2016; Thelwall, in press-a, submitted; Hug & Brändle, 2017) but allows automatic data harvesting. It is therefore a promising source of citation data for large scale citation analyses. It should be especially useful for fields with many online publications and for recently-published research since it includes citations from preprints (Thelwall, in press-a, submitted). Nevertheless, one important limitation is that it does not allow DOI searches (Hug, Ochsner, & Brändle, 2017) and so it is not clear whether it is possible to obtain reasonably comprehensive sets of Microsoft

---

[1] Thelwall, M. (2018). Microsoft Academic automatic document searches: accuracy for journal articles and suitability for citation analysis. Journal of Informetrics, 12(1), 1-7. doi:10.1016/j.joi.2017.11.001

Academic citation counts for individual journal articles. This would be necessary for citation analysis applications that focus on documents rather than researchers, journals or fields.

Microsoft Academic uses information from publishers and the Bing web search engine to create a database of documents that it classifies as academic-related. Academic fields, authors, institutions, documents, venues (journals and conference series), and events (individual conferences) are classified into 'entities', that are related to each other in a graph-like structure (Sinha, Shen, Song, Ma, Eide, Hsu, & Wang, 2015). When a new paper is found, Microsoft Academic attempts to extract these six types of information from it using publisher information and data mining techniques. For example, paper authors are conflated with existing authors from other documents if they have the same name and email address/institution, or if they obey other author name disambiguation rules. Authors that cannot be matched with any previously found are assigned to a new author entity. Similarly, new document records are matched with a previously found documents if the metadata is close enough. Allowances are made for papers with the same title to be mentioned in different ways, in a process called "title conflation" (Sinha et al., 2015). The heuristics and exploitation of web data have combined to introduce errors into some Microsoft Academic records (Harzing, 2016; Harzing, & Alakangas, 2017a; Hug, Ochsner, & Brändle, 2017) but these have decreased over time (Harzing, & Alakangas, 2017b).

Although it is useful to correlate any new source of citations with Scopus or Web of Science citation counts (Sud & Thelwall, 2016), most investigations into Microsoft Academic have focused instead on the relative magnitude of the two or on changes in the ranking of academics or groups of publications if one source is used rather than the other (Harzing, 2016; Harzing, & Alakangas, 2017a; Hug, Ochsner, & Brändle, 2017). An investigation of the early trial version compared rankings of journals based on Microsoft Academic citations and Scopus citation counts from Scimago (presumably per paper in both cases), finding a correlation of 0.9 (Herrmannova & Knoth, 2016). An analysis of 29 large monodisciplinary journals from different fields found article-level correlations for journals between 0.89 and 0.98 with Scopus citations (Thelwall, in press). Microsoft Academic seems to be unable to extract references from a minority (11%) of documents that it has full text access to, although the reason for this is unclear (Haunschild, Hug, Brändle, & Bornmann, 2017).

The success of Scopus and WoS in extracting cited references (Franceschini, Maisano, & Mastrogiacomo, 2013) gives information that is relevant to the task of matching queries to documents. Journal or publisher referencing formats (Franceschini, Maisano, & Mastrogiacomo, 2014), references overlooked by the citation index (Franceschini, Maisano, & Mastrogiacomo, 2015a) and author typographic errors in reference lists (Franceschini, Maisano, & Mastrogiacomo, 2016ab) can create problems specific to reference linking. The presence of errors in the bibliographic records for articles in Scopus and WoS affects the effectiveness of the queries used in the current paper. Most importantly, DOIs are not always correct in WoS and Scopus (Franceschini, Maisano, & Mastrogiacomo, 2015b). Similarly, titles and author names are sometimes incorrect in references, presumably due to scanning errors (for 0.1% of references in Scopus: Franceschini, Maisano, & Mastrogiacomo, 2015b), and similar occasional errors may also occur for bibliographic records. The task of matching published articles with online first versions is also error-prone (Franceschini, Maisano, & Mastrogiacomo, 2016ab). Scopus and WoS seem to occasionally correct mistakes in old articles (Franceschini, Maisano, & Mastrogiacomo, 2016c) and so the error rate may vary over time.

The Microsoft Academic Applications Programming Interface (API) allows a limited number of free automatic queries per month, with the possibility to pay for more. It supports queries for a limited set of information recorded by Microsoft Academic. Queries can either be precise or approximate. Precise queries request information about an entity (e.g., author, journal, paper) by referencing it with its official Microsoft Academic entity identifier or its official Microsoft Academic (lower case) normalised name. For example, the precise author query *jun li* returns information about University of Michigan Associate Professor of Human Genetics Jun Li (author ID: 2552318488) rather than University of Michigan Assistant Professor of Technology and Operations Jun Li (author ID: 2682221189) or any other Jun Li. Individual papers can be searched for by title ID or exact lower case normalised entity name, otherwise approximate title search matches will be returned. Although less common than for author names, different journal articles can have the same title (e.g., at least four, "The future of our profession"). Incorrect matches can be narrowed down by adding the publication year, author names or journal name but not DOI.

Since Microsoft Academic uses data mining heuristics to gather information about documents and, as discussed above, document matching is error prone, it is not clear how often document records can be identified. It is possible to identify documents through Microsoft Academic title queries either by searching for the title string (precision 0.897, recall 0.512 for University of Zurich repository documents) or for the individual words in the title (precision 0.707, recall 0.514 for University of Zurich repository documents) (Hug & Brändle, 2017). These figures include identifying the best matches by comparing DOIs, when present, titles, or other bibliographic information (in that order). Other approaches have not been assessed and methods have not been systematically tested on different fields. In addition, the recall figure seems too low to be useful but recall may be higher for data sets consisting of just journal articles. This article addresses this issue, with the results used to analyse the correlation between Microsoft Academic and Scopus citations for journal articles across more fields than ever before.

## 2   Research questions

This article proposes and assesses the effectiveness of a range of methods to identify journal articles in Microsoft Academic. The objective is to provide recommendations for the best methods to use alongside estimates of their performance. A secondary goal is to use the results to assess the correlation between citations from the two sources to assess whether Microsoft Academic document searches could be used to get data for citation analysis.

- RQ1: What is the optimal method to identify matching journal articles with DOIs in Microsoft Academic and how accurate is it?
- RQ2: What is the optimal method to identify matching journal articles *without* DOIs in Microsoft Academic and how accurate is it?
- RQ3: Are there any fields in which Microsoft Academic citation counts might reflect a different kind of impact to Scopus citation counts?

## 3   Methods

The research design was to obtain a large sample of relatively recent, but not current, research articles and to check the accuracy of Microsoft Academic queries for them using a range of plausible query types. The core analysis strategy was to use DOIs as unique document identifiers to check for the accuracy of the search results.

## 3.1 Data

The article sample was drawn from Scopus-indexed journal articles published in 2012. There is no comprehensive index of journal articles and so Scopus was chosen as the largest international citation index used for research evaluation purposes (Moed, & Visser, 2008; Mongeon & Paul-Hus, 2016). The year 2012 was selected for being old enough to be extensively covered online (five years before data collection), whilst still being relatively recent. Only documents of the Scopus type "Journal Article" were included since this is the type primarily used in research evaluations.

Scopus categorises academic research into 335 sub-fields (e.g., 1111 Soil Science), spanning almost all areas of academia (although not equally). Articles are categorised by their publishing journal and many journals are assigned to multiple sub-fields. The last up to 5,000 articles from 2012 in each of the 335 sub-fields was downloaded in August 2017. The number 5,000 is a Scopus system restriction. Seven sub-fields had no articles in 2012, leaving 328 subfields with 1,257,232 articles. Articles without DOIs were discarded since DOIs were needed for the accuracy checks. This eliminated two more fields, giving 326 fields with 1,005,074 (79.9%) articles with DOIs.

To limit the cost of the queries, a random sample of 400 articles (without replacement) was taken from each set (using the dot net random number generator function RND()), giving 126,400 articles with DOIs from the 326 fields. The number 400 was chosen heuristically to give large enough samples to detect large differences between subfields without giving an unnecessary degree of extra accuracy. Three fields with under 50 articles were removed (13, 13, and 44) since these could give inaccurate results, giving a final set of 323 fields and 126,312 journal articles with DOIs.

This sample probably includes nearly all academic fields that publish in journals but is likely to be somewhat biased. For example, fields in which scholars rarely publish in journals, or for which Scopus rarely indexes the journals, are under-represented. In addition, subfields vary in size from hundreds to tens of thousands of documents per year. Nevertheless, the final sample has wide coverage of academia and is broadly representative of Scopus coverage.

## 3.2 Document query construction

Microsoft academic queries were built for each document using four plausible strategies and, for comparison purposes, a more comprehensive version. All searches include article titles since this seems to be a basic requirement, but authors, journal, and publication year could also be queried. Only the first author is included in the author searches because, as the Jun Li example above illustrates, author name disambiguation seems to be a difficult task and so including multiple authors seems likely to substantially degrade query performance. The details of the searches were designed during a period of trial and error with a small set of test queries from one field. Approximate rather than precise queries were used in all cases because normalised names were not known. To avoid over-fitting, the methods were applied to the full set of 323 fields only once and were not modified in response to issues found.

***Author names***: Author name queries used the first initial and last name in an approximate author name search. All characters were converted to lower case and accented characters were converted to non-accented equivalents. Hyphens and apostrophes were replaced with spaces. For example, the Scopus author name *Jehlička J.* was converted to the query, **Composite(AA.AuN='j jehlicka')**. Full first names were not used since these may cause more correct matches to be rejected. This process may be ineffective for authors with double last names, as is common in Spain (e.g., Juan Manuel Ayllón Millán), which may reduce the matching accuracy and introduce international bias into the results.

*Journal names*: Journal name queries used the full journal name, converted to lower case. Accented characters were converted to non-accented equivalents. Ampersands were replaced by the term 'and'. Non alpha-numeric characters, such as colons, were replaced with spaces. For example, the journal *Agris On-line Papers in Economics and Informatics* was converted to, **Composite(J.JN='agris on line papers in economics and informatics')**.

*Publication year*: The Scopus publication date was truncated to the publication year, which was always 2012, giving the query, **Y=2012**.

*Title*: All characters were converted to lower case and accented characters converted to non-accented equivalents. Tags in titles for subscripts or superscripts (e.g., <sup>, </inf>) were removed. Greek letters were replaced with their names in words (e.g., α to alpha). Non alpha-numeric characters were replaced with spaces. For example, *The O/OREOS mission: First science data from the space environment viability of organics (SEVO) payload* was converted to the query, **Ti='the o oreos mission first science data from the space environment viability of organics sevo payload'**.

*Combinations*. When multiple document parts were included in a query they were combined with the And operator, as in the following example, **And(Composite(AA.AuN='c lin'),Composite(J.JN='biometrika'), Ti='designs of variable resolution', Y=2012)**.

The queries were submitted to the Microsoft Academic API between 2 and 11 September 2017 using a free trial key.

### 3.3 Analysis

For each query, the Scopus-recorded DOI for the intended document was checked against the Microsoft Academic-recorded DOI and the query result was classified as correct if the two matched, after converting both to lower case and filtering out trailing dots.

For the matching process that ignored DOIs, the query document metadata was compared against the result document metadata and the result was rejected if the two were too different. Small variations were allowed, given that errors could be introduced in either source. The process allowed differences in one of: title, publication year, author, journal name. Cases with at least two differences were rejected. Articles were also rejected if there was an overlap of less than 85% in the words used in their titles. These rules were worked out on a pilot set from a single discipline and could probably be improved with a larger study.

The following standard information retrieval statistics were calculated for each sub-field and query type.
- *Precision*: The percentage of documents returned by the query with matching DOIs.
- *Recall*: The proportion of queries that had at least one result with a matching DOI.

For the third research question, Spearman correlations were calculated between citation counts from the two sources for each field. Actual (linked) counts rather than estimated citation counts were used from Microsoft Academic. Correlations with Scopus or WoS citation counts are the logical first tests for any new indicator (Sud & Thelwall, 2014). Spearman correlations are preferable to Pearson values because citation data is skewed.

The average values of the citation counts were also compared to detect whether the relative magnitude of the two citation counts varied by field. Geometric means were used instead of arithmetic means, also due to skewed data (Thelwall & Fairclough, 2015; Zitt, 2012).

## 4 Results

For the first research question, the objective is to identify as many matching journal articles in Microsoft Academic as possible, using DOIs to remove false matches. Thus, the optimal query type is the one that obtains the most correct matches (greatest recall). From Table 1, this is a title-only query, with a recall of 89.6% on average. Thus, the optimal strategy (of the five tested) to search Microsoft Academic to find a matching document is to query the document title and then remove all matches with incorrect DOIs. This returns, on average, a match for 89.6% of all documents for any field. The worst-case is that this method finds only 68.5% matching documents in a field and the best case is that it finds a match for 99.3%. Thus, adding either the year, journal name or author to the query removes some correct matches and does not add new correct matches.

Full results for all tables are available in the online Appendix spreadsheet.

Table 1. DOI checks of whether documents returned by Microsoft Academic are correct matches of the original documents generating the query (n=323 fields from 2012, n= 126,312 journal articles, 76 to 400 articles per field; mean 391.1 articles per field).

|      | Full query | | Author, title | | Journal, title | | Year, title | | Title | |
|------|--------|-------|--------|-------|--------|-------|--------|-------|--------|-------|
|      | Recall | Prec. | Recall | Prec. | Recall | Prec. | Recall | Prec. | Recall | Prec. |
| Min. | 19.0%  | 73.4% | 63.0%  | 72.9% | 19.3%  | 69.4% | 68.2%  | 73.8% | 68.5%  | 65.7% |
| Max. | 94.1%  | 100%  | 96.6%  | 100%  | 96.9%  | 100%  | 99.3%  | 100%  | 99.3%  | 99.7% |
| Med. | 71.0%  | 99.4% | 86.0%  | 97.5% | 75.5%  | 99.0% | 90.0%  | 97.6% | 90.8%  | 95.7% |
| Mean | 69.8%  | 98.8% | 84.4%  | 96.2% | 74.2%  | 97.6% | 88.6%  | 96.4% | 89.6%  | 93.8% |

For the second research question, documents with DOIs are used to test the accuracy of queries for documents without DOIs. From Table 2, the optimal query is the same as for documents with DOIs. More specifically, the method to obtain the largest number of matching documents without DOIs in Microsoft Academic is to search for the document titles then remove matches for which at least two of the publication year, author, or journal are different or the title is too dissimilar. This will return, on average 89.1% matches (i.e., 0.5% less than for DOI matching). The worst-case is to have 67.0% matches and the best case is 100% matches. Although it is possible that matching articles with similar authors, journals and publication years are different articles, this can be checked by matching DOIs (Table 3). From the precision column in Table 3, checking authors, journals and publication years is enough to filter out incorrect matches 98.7% of the time, with the worst case being 87.1%.

Table 2. Metadata checks of whether documents returned by Microsoft Academic match the original documents generating the query (n=323 fields from 2012, n= 126,312 journal articles, 76 to 400 articles per field; mean 391.1 articles per field).

|      | Full query | | Author, title | | Journal, title | | Year, title | | Title | |
| --- | --- | --- | --- | --- | --- | --- | --- | --- | --- | --- |
|      | Recall | Prec. | Recall | Prec. | Recall | Prec. | Recall | Prec. | Recall | Prec. |
| Min. | 20.3% | 73.9% | 62.5% | 73.1% | 20.3% | 68.7% | 67.0% | 74.2% | 67.0% | 65.2% |
| Max. | 92.9% | 100% | 96.6% | 100% | 96.9% | 100% | 100% | 99.7% | 100% | 99.2% |
| Med. | 71.5% | 99.7% | 86.7% | 98.1% | 75.5% | 98.8% | 89.8% | 97.2% | 90.5% | 94.7% |
| Mean | 70.3% | 99.6% | 85.1% | 96.9% | 74.2% | 97.7% | 88.3% | 96.1% | 89.1% | 93.3% |

Table 3. DOI checks of whether the documents classed as matching following the metadata checks of Table 2 also have matching DOIs (n=323 fields from 2012, n= 126,312 journal articles, 76 to 400 articles per field; mean 391.1 articles per field).

|      | Full query | | Author, title | | Journal, title | | Year, title | | Title | |
| --- | --- | --- | --- | --- | --- | --- | --- | --- | --- | --- |
|      | Recall | Prec. | Recall | Prec. | Recall | Prec. | Recall | Prec. | Recall | Prec. |
| Min. | 18.5% | 84.4% | 62.5% | 87.0% | 18.5% | 83.9% | 67.0% | 87.1% | 67.0% | 87.1% |
| Max. | 92.9% | 100% | 96.6% | 100% | 96.4% | 100% | 99.3% | 100% | 99.3% | 100% |
| Med. | 70.8% | 99.5% | 85.5% | 99.2% | 74.8% | 99.4% | 88.5% | 99.2% | 89.3% | 99.2% |
| Mean | 69.6% | 98.9% | 84.0% | 98.6% | 73.4% | 98.8% | 87.2% | 98.8% | 87.9% | 98.7% |

## 4.1 Reasons for queries returning incorrect or no matches

There are many different reasons for apparent or actual incorrect matches in the data. These are discussed here to check the results and give them context.

The Microsoft Academic records returned by the queries did not all have DOIs. On average, 98.2% had a DOI, with percentages for individual fields varying from 80.0% (Accounting) to 100%. This small percentage is not enough to explain the 89.6% recall for the optimal title search method (DOI checking). Thus, most of the articles without any matching records returned by Microsoft Academic either were not in the Microsoft Academic database or were indexed in such a way that a title search did not find them. Missing DOIs explained the lowest recall category (Accounting) in Table 1, however.

Preliminary testing had found that both Scopus and Microsoft Academic records contained some errors in author names, journal names and publication years (as previously found for Scopus: Franceschini, Maisano, & Mastrogiacomo, 2015b), which accounts for the higher recall for the title-only searches despite using approximate match queries (single equals signs in the queries). Some title differences are also likely between Microsoft Academic and Scopus because titles are not always recorded consistently. For example, authors and citers may shorten titles, omit punctuation, omit initial articles, and display formulae, superscripts, subscripts, unusual letters and accents differently (see also: Franceschini, Maisano, & Mastrogiacomo, 2015a; Franceschini, Maisano, & Mastrogiacomo, 2016ab).

Missing records for **Critical Care and Intensive Care Medicine** were investigated to find reasons why a category could have a lower than average percentage of matches for Microsoft Academic document searches. This category was chosen for a low percentage of matching articles (81.7%) and a higher percentage of Microsoft Academic articles with DOIs (91.4%). Critical Care and Intensive Care Medicine contained 38 Scopus records without Microsoft Academic matches. The missing results included all articles in Scopus from *Enfermeria Intensiva* (3), *Notarzt* (4), *Revista Brasileira de Terapia Intensiva* (9), *Revista Espanola de Anestesiologia y Reanimacion* (3), *Anasthesiologie Intensivmedizin Notfallmedizin Schmerztherapie* (7), *Chinese Critical Care Medicine* (2), and *Intensiv- und Notfallbehandlung* (1). Thus, entire years of journals missing from Microsoft Academic explains 29 of the missing 38 articles. In some cases, the article titles were dual language and indexed in different languages in Scopus and Microsoft Academic (*Notarzt, Revista Brasileira de Terapia Intensiva*). In other cases, journal articles were apparently inconsistently indexed in different languages in Microsoft Academic (*Anasthesiologie Intensivmedizin Notfallmedizin Schmerztherapie*) or the journals did not seem to be indexed (*Enfermeria Intensiva, Revista Espanola de Anestesiologia y Reanimacion, Chinese Critical Care Medicine, Intensiv- und Notfallbehandlung*). The remaining 9 articles were checked for accuracy via DOI searches and then searched for in Microsoft Academic using different title variants, with the following results.

- "Asthma phenotypes" by Pavord could not be found in Microsoft Academic.
- "Acute myocardial infarction and cardiogenic shock: Prognostic impact of cytokines: INF-γ, TNF-α, MIP-1β, G-CSF, and MCP-1β" was found in Microsoft Academic with the record "acute myocardial infarction and cardiogenic shock prognostic impact of cytokines inf γ tnf α mip 1β g csf and mcp 1β". Webometric Analyst's title search was, Ti='acute myocardial infarction and cardiogenic shock prognostic impact of cytokines inf γ tnf alpha mip 1beta g csf and mcp 1beta' and so in this case, the conversion of β into beta for the query (based on previous experiments) had not worked. The same problem occurred for, "Previous prescription of β-blockers is associated with reduced mortality among patients hospitalized in intensive care units for sepsis". This is therefore a Webometric Analyst query generation error.
- "Neuro intensive" and "Infectious diseases - Pathogen epidemiology, the development of resistance", are both from the same journal and without authors. They do not seem to be peer reviewed academic articles and are therefore Scopus document type classification errors.
- "Echocardiography in emergency admissions: Recognition of cardiac low-output failure" was recorded in Microsoft Academic with a different title, combining its German and English titles, "Echokardiographie in der Notaufnahme@@@Echocardiography in emergency admissions: Erkennen des kardialen Low-Output-Versagens@@@Recognition of cardiac low-output failure". This is a Microsoft Academic indexing error.
- "Splenic tuberculosis in HIV patient" was recorded in Microsoft Academic with its (primary) Spanish name, "Tuberculosis esplénica en paciente con VIH". This is therefore a dual language issue.
- "Validation of the Better Care® system to detect ineffective efforts during expiration in mechanically ventilated patients: A pilot study" was not indexed in Microsoft Academic but its erratum was, "Erratum to: Validation of the Better Care® system to detect ineffective efforts during expiration in mechanically ventilated patients: a pilot study". The Microsoft Academic title conflation policy thus equated an erratum with the original article.
- "Patient recovery and the post-anaesthesia care unit (PACU)" was found by a search in the Microsoft Academic web interface, so the reason why it was not found in the API search is unknown.

In summary, whilst the main reason for articles not being found is that they are from journals not indexed by Microsoft Academic, language issues, document type misclassifications, errata and query errors were all minor contributory factors.

Another category, **Materials Science (all)** with 97.3% Microsoft Academic DOIs but 81% recall was also investigated. For this category, all 17 *Acta Crystallographica Section E: Structure Reports Online* articles were not found in Microsoft Academic, perhaps because their online nature failed an indexing test. All 7 *Zeitschrift fur Kristallographie - New Crystal Structures* articles were also not found but in this case they were present but the queries had not worked. The titles were all complex, indicating a likely Webometric Analyst query construction failure. For example, "Crystal structure of 2-hydroxybenzylidene-5-methylisoxazole-4-carbohydrazide monohydrate, C$_{12}$H$_{13}$N$_3$O $_4$" had been translated by Webometric Analyst into the title query, Ti='crystal structure of 2 hydroxybenzylidene 5 methylisoxazole 4 carbohydrazide monohydrate c12h13n3o 4' whereas the document name in Microsoft Academic did not have a space before the final digit, despite the space appearing in the Scopus version of the title. This space is not in the original article title so its presence is apparently a Scopus indexing error. Thus, missing unusual journal issues and complex titles accounted for the missing articles in this case.

A final category checked, **Classics**, also revealed journal-level issues. Most or all articles in several journals had not been found: *Ancient Near Eastern Studies* (7 out of 8 missing); *Babesch* (4 out of 6); *Emerita, Revista de linguistica y filologia clasica* (13 out of 15); *Kadmos* (9 out of 16); *Klio* (9 out of 12), *Philologus* (18 out of 22); *Revue des Etudes Grecques* (all 7); *Zeitschrift fur Antikes Christentum* (all 2). Thus, journal-level indexing issues seem to be the biggest problem overall.

From a scientometric perspective, the language issues and problems with some non-English journals are the most serious problem because they can produce international biases in the results. For example, if UK and Spanish research productivity is compared through matches, then Spanish research seems likely to be less represented and so corrective action would need to be taken. To check further for language issues, the percentage of (title) matches found for articles were checked for all 323 fields based on the affiliation country of the first author. Articles by authors from English-speaking countries are more likely to have their articles retrieved by Microsoft Academic queries (Table 4), with the difference being up to 18% (Australia: 95%; Brazil: 77%). Countries that publish predominantly in English tend to have a higher proportion of matches (the last two columns of Table 4 have a correlation of 0.74), irrespective of their national language (e.g., Taiwan). India is a partial exception.

Table 4. The percentage of Microsoft Academic matches found from title searches, by first author country affiliation in Scopus (n=323 fields from 2012, n= 126,312 journal articles) for the 25 most common affiliations. The final column reports the percentage of Scopus articles with the country affiliation with a registered language of English.

| Country* | Articles | Matches in MA | Percentage matches | Percentage English in Scopus** |
|---|---:|---:|---:|---:|
| **Australia** | 3610 | 3447 | 95% | 100% |
| Norway | 792 | 750 | 95% | 96% |
| Sweden | 1515 | 1424 | 94% | 100% |
| **United States** | 29777 | 27898 | 94% | 100% |
| Israel | 896 | 839 | 94% | 100% |
| **United Kingdom** | 7496 | 7012 | 94% | 100% |
| Russian Federation | 1458 | 1363 | 93% | 92% |
| **Canada** | 4217 | 3940 | 93% | 99% |
| Taiwan | 1964 | 1831 | 93% | 99% |
| Netherlands | 2290 | 2134 | 93% | 97% |
| Denmark | 859 | 799 | 93% | 96% |
| Turkey | 1652 | 1518 | 92% | 94% |
| Iran | 1737 | 1594 | 92% | 97% |
| Italy | 3658 | 3352 | 92% | 96% |
| South Korea | 3297 | 3008 | 91% | 97% |
| India | 4004 | 3652 | 91% | 100% |
| Japan | 5106 | 4599 | 90% | 93% |
| Belgium | 1094 | 985 | 90% | 95% |
| Poland | 1234 | 1101 | 89% | 91% |
| Switzerland | 1210 | 1079 | 89% | 95% |
| China | 14270 | 12290 | 86% | 79% |
| Germany | 5438 | 4639 | 85% | 90% |
| Spain | 3613 | 2963 | 82% | 88% |
| France | 4177 | 3245 | 78% | 87% |
| Brazil | 3139 | 2411 | 77% | 90% |

*Bold countries are English-speaking.
**E.g., for Australia, divide the results of the Scopus query *PUBYEAR = 2012 AND AFFILCOUNTRY(Australia) AND LANGUAGE(English)* by the results of the Scopus query *PUBYEAR = 2012 AND AFFILCOUNTRY(Australia)*.

## *4.2 Correlations with Scopus citations and average citation counts*

The citation counts from Scopus and Microsoft Academic are similar overall in average magnitude (6.74 vs. 6.85) and in terms of correlation (0.948) (Table 5). Nevertheless, there are some fields in which the two differ more substantially.

*Assessment and Diagnosis* (n=166) has a correlation of 0.951 but unusually many citations from Scopus (5.96) compared to Microsoft Academic (2.65). This seemed to be due to Microsoft Academic not indexing the very large *Nursing* magazine in this category – and perhaps other health journals or magazines in related categories.

*History and Philosophy of Science* (n=224) has a correlation of 0.875 but unusually few Scopus citations (2.97) compared to Microsoft Academic citations (5.18). This is presumably due to Microsoft Academic indexing more non-article sources than Scopus, such as edited book series and theses.

*Classics* (n=220) has the lowest correlation of 0.625. This low correlation is due to many zeros in the data set (for an explanation, see: Thelwall, 2016), with few Scopus citations (0.89) and Microsoft Academic citations (0.44).

Table 5. Spearman correlations and geometric means for articles passing the DOI match test in Table 2 (n=323 fields, n= 126,312 journal articles).

|         | Articles | Scopus citations geomean | Micro. Acad. citations geomean | Difference (% of MA) | Spearman correlation |
|---------|----------|--------------------------|--------------------------------|----------------------|----------------------|
| Min     | 57       | 0.73                     | 0.44                           | -3.31 (-124.9%)      | 0.625                |
| Max     | 357      | 22.14                    | 21.54                          | 3.03 (42.7%)         | 0.995                |
| Median  | 280      | 6.53                     | 6.65                           | 0.13 (2.3%)          | 0.958                |
| Average | 272.2    | 6.74                     | 6.85                           | 0.11 (0.8%)          | 0.948                |

## 5 Discussion

The main limitations of the experiment include that query recall may be lower for much older articles if they are less represented online or for very recent articles, which may be less likely to be found by Microsoft Academic. Indexing errors in Scopus (Franceschini, Maisano, & Mastrogiacomo, 2015b, 2016b) also undermine the results and these vary for different publication years (Franceschini, Maisano, & Mastrogiacomo, 2016c). In addition, recall may be lower for articles not indexed in Scopus and for articles without DOIs if these are more likely to be in journals that have little publicly available information. There may also be some types of articles with repetitive titles or short titles that Microsoft Academic does not index well because of title conflation. The recall figures are minimum bounds for what is possible and alternative query strategies may produce slightly higher results, but this is unlikely to affect the overall conclusions. The results are partly dependant on the matching process in Microsoft Academic and partly on its ability to find documents (Harzing, & Alakangas, 2017b), both of which may change over time. Finally, although Microsoft Academic has been formally released after over a year as a trial (beta) version, it may evolve to render the results obsolete. If it introduces a DOI search, then the first research question would be redundant.

The optimal method is the same as the one used in a previous paper (Hug & Brändle, 2017) but it has substantially higher precision and recall for sets of journal articles than previously reported for University of Zurich repository documents.

The high or very high correlation magnitudes show that Microsoft Academic and Scopus citation counts are essentially interchangeable for articles with DOIs. Since correlations are affected by the magnitude of citation counts (Thelwall, 2016) they can be expected to be lower for more recent articles even if the underlying relationship is strong. The correlations are within the range for monodisciplinary journal-level correlations (0.89 to 0.98) with Scopus citations (Thelwall, in press) and for five broad fields based on the publications 2008-2015 of a single university (Hug & Brändle, 2017). The current paper extends these results by showing that individual journals can lower correlations, that title language is an issue, and that low citation fields can have lower correlations. Nevertheless, the results confirm that there is overall a strong relationship between Scopus and Microsoft Academic citation counts.

There are national differences of up to 18% in the completeness of the results (Table 4) that seem to be mainly due to problems generating effective queries for many languages other than English. These could be from problems matching or scanning accented or non-ASCII characters or from confusion over articles with dual language titles. The reduced accuracy may also reflect cultural differences in names (e.g., Spanish double last names), or under-resourced national journals that have non-automated document management procedures and are less able to ensure consistent and accurate indexing in all sources.

There are three patterns in the disciplines with lower correlations between Scopus and Microsoft Academic citation counts. Fields with lower overall citation counts tend to have lower correlations between the two sources (Pearson correlation of 0.59 between geometric mean Scopus citation counts and Scopus/Microsoft Spearman correlations, n=323). This could either be because the relationship is weaker in low citation fields or because the low number masks the underlying relationship strength (Thelwall, 2016). A second pattern is that the correlations tend to be low for Arts and Humanities fields (below 0.9 for 9 out of 14 and below 0.8 for 3). Four of the Arts and Humanities exceptions have strong elements of science or social science (Language and Linguistics; Archeology (arts and humanities); Conservation; Music). Four of the 15 mathematics related subjects also have correlations below 0.9 (Mathematics (all); Algebra and Number Theory; Discrete Mathematics and Combinatorics; Geometry and Topology).

Combining the two parts of this study, given that it is possible to query for articles reasonably accurately and comprehensively in Microsoft Academic using the methods introduced in this article and that the resulting Microsoft Academic citation counts have high correlations with Scopus citation counts, it is possible for evaluators with sets of articles to analyse to use Microsoft Academic citations as a substitute for Scopus (or WoS) citations. This may be desirable for researchers without access to Scopus or WoS.

In comparison to the Scopus API, Microsoft Academic automatic queries are probably cheaper for users that do not already have access to Scopus, depending on the task. At the time or writing the cost was €0.22 per 1,000 transactions[2], with each document search consuming one transaction. The cost of an institutional subscription to Scopus[3] is not fixed and so a direct comparison is not possible. The cheaper option (Microsoft Academic searches or Scopus with API use) will depend on the size and geographic location of the host institution and the number of queries to be submitted.

## 6 Conclusions

The results show that it is possible to search for journal articles from almost all fields in Microsoft Academic with a high degree of precision and recall, irrespective of whether the source document has a DOI. The optimal title search method is one previously proposed for all documents (Hug & Brändle, 2017). Microsoft Academic is therefore a practical tool for evaluating the impact of a set of journal articles. Incomplete coverage may be a problem for groups that publish primarily in journals with dual language titles, but this problem could presumably be solved by searching in both languages. Microsoft Academic's advantage for citation analysis would be higher citation counts for recently-published articles (Harzing, & Alakangas, 2017a) and the ability to include articles not indexed by Scopus (Hug & Brändle, 2017; Thelwall, submitted-b). Caution should be exercised when using the results to compare national contributions, however, since there are substantial national differences in the ability of Microsoft Academic to find articles, at least using the methods described here.

---

[2] https://azure.microsoft.com/en-gb/pricing/details/cognitive-services/academic-knowledge-api/
[3] https://www.elsevier.com/solutions/scopus/support/activating-scopus

The almost universally high correlations between Scopus and Microsoft Academic citations confirm previous research for journals and authors (Harzing & Alakangas, 2017b; Thelwall, in press) and demonstrate that there are only minor anomalies, at least for articles from 2012. These anomalies are partly related to the indexing of dual language article titles and so if this occurs for a study then articles should be searched for twice, one in each language.

To support others to conduct document queries, the code used in the current paper has been added to the free software Webometric Analyst (http://lexiurl.wlv.ac.uk in the Citations menu). This includes a program to generate queries from document metadata, to submit the queries to Microsoft Academic and save the results, and to filter out false matches using DOIs, if present, and/or metadata. Publish or Perish (https://harzing.com/resources/publish-or-perish) can also be used to submit Microsoft Academic API queries and works on non-Windows platforms but does not yet have document search and filtering capabilities.

A practical limitation with Microsoft Academic is that it costs money to submit more than 1000 queries per month, so large studies may be expensive. This should not be a problem for research evaluations of individual departments and it is likely to be cheaper than purchasing a WoS or Scopus subscription (for institutions that do not have them), depending on the study. Nevertheless, if additional articles are needed for comparison or field-normalisation purposes (Thelwall, 2017; Waltman, van Eck, van Leeuwen, Visser, & van Raan, 2011), then this may trigger charges. Another practical limitation is that, like Google Scholar (Delgado López-Cózar, Robinson-García, & Torres-Salinas, 2014; Kousha & Thelwall, 2008), citation counts in Microsoft Academic can be manipulated easily by adding lower quality citation-rich documents to the web or archives that it indexes (e.g., SSRN: Thelwall, in press). It should not therefore be used in formal evaluations where the participants are told in advance (Wouters & Costas, 2012), such as the UK REF, because the results could be gamed.